# Investigation of radiation hardness of silicon semiconductor detectors under irradiation with fission products of $^{252}$Cf nuclide.

N V Bazlov[1,2], A V Derbin[1], I S Drachnev[1], I M Kotina[1], O I Konkov[1,3], I S Lomskaya[1], M S Mikulich[1], V N Muratova[1], D A Semenov[1], M V Trushin[1] and E V Unzhakov[1]

[1] NRC "Kurchatov Institute" - PNPI, Gatchina, Russia
[2] Saint-Petersburg State University, Universitetskaya nab. 7/9, St. Petersburg, Russia
[3] Ioffe Physical-Technical Institute of the Russian Academy of Sciences, St. Petersburg, Russia

e-mail: trushin_mv@pnpi.nrcki.ru

**Abstract**. Influence of the prolonged irradiation by fission products of $^{252}$Cf radionuclide on the operational parameters of silicon-lithium Si(Li) p-i-n detectors, Si surface barrier detectors and Si planar p$^+$n detector was investigated. The obtained results revealed a linear shift of the fission fragment peaks positions towards the lower energies with increase of the irradiation dose for all investigated detectors. The rate of the peaks shift was found to depend strongly on the detector type and the strength of the electric field in the detector's active region, but not on the temperature of irradiation (room or liquid nitrogen temperature). Based on the obtained results, the possibility of integration of the investigated types of Si semiconductor detectors in a radionuclide neutron calibration source is considered.

## 1. Introduction

Heavy nuclides subjected to spontaneous fission decay accompanied by emission of several fast neutrons can be utilized as a compact neutron calibration source. The most common spontaneous fission source is $^{252}$Cf which undergoes α-decay and spontaneous fission with a branching ratio of 97:3, whereas each spontaneous fission event liberates 3.8 neutrons and 9.7 gamma-ray photons on average [1]. The timing of the moment of neutron production can be fixed by detecting the fission fragments signal with a semiconductor detector.

Semiconductor detectors possess sufficiently high energy resolution for detection of the high-energy heavy ions. The main obstacle for the integration of such detectors in the neutron calibration source could be their limited lifetime under the influence of the nuclide radiation [2]. Degradation of the detector's operational parameters effectively proceeds just in case of irradiation by alpha particles and fission fragments (FF), which are capable of transferring a significant fraction of their energy to the atoms of the detector lattice. Therefore, the degradation of the semiconductor detector will limit the maximum neutron source activity and/or the source expiration period.

This article is devoted to the investigations of degradation of the operational parameters of several types of silicon semiconductor detectors under prolonged irradiation with fission products of $^{252}$Cf (α-particles and fission fragments). The main issue was to study the rate of degradation of different detector types under irradiation by $^{252}$Cf fission products at various irradiation conditions. Irradiation was performed at room and liquid nitrogen temperatures as well as with different detector's operational biases, i.e. with different electric field strength in the detectors active regions. Results of the preceding investigations were presented in previous articles [3-5].

## 2. Detectors and experimental setup

Three types of silicon semiconductor detectors were under investigations. Detectors of the first type are SiLi *p-i-n* detectors produced from p-type silicon ingot with resistivity of 2.5 kΩ×cm and carrier lifetime of 1000 µs. Two similar detectors with a sensitive region of 20 mm in diameter and 4 mm thick were produced using standard Li drift technology [6]. The thickness of the undrifted p-type layer in these detectors (i.e. the entrance window thickness) usually amounts to 300-500 nm [7], which is kept to suppress the excessive growth of the leakage current at high operation reverse voltage [8]. Detectors of the second type were two surface-barrier (SB) detectors fabricated from p-type boron-doped silicon wafer of (111) orientation and 10 mm in diameter. The resistivity and the carrier lifetime were 1 kΩ×cm and 1000 µs, respectively. The front side of the wafers was covered by a thin layer of amorphous silicon which served as a passivation coating [9]. The ohmic contact was made by sputtering of Pd layer on the whole rear side of the wafer, whereas the rectifying one – by evaporation of Al dot with diameter of 7 mm in the center of the wafer's front side. Detector of the third type was $p^+n$ planar detector with the thickness of 300 µm produced in Ioffe Physical-Technical Institute (entrance window thickness was about 50 nm and the voltage of full depletion – nearly 150 V).

Irradiation by a $^{252}$Cf source was performed in vacuum cryostat typically during 10-20 days. The source representing a stainless steel substrate covered by an active layer under the thin protective coating was mounted 1 cm above the detector front surface that was collimated in order to exclude side surface effects of incomplete charge collection. The spectra of the fission products of $^{252}$Cf were recorded continuously during the whole irradiation period in short 1-hour series, what allowed us to observe the spectra evolution directly. Detector reverse current was also monitored during the whole irradiation period on 5-second basis with the following averaging on 1-hour measurement series. Details of the measurement setup were presented in [3-5].

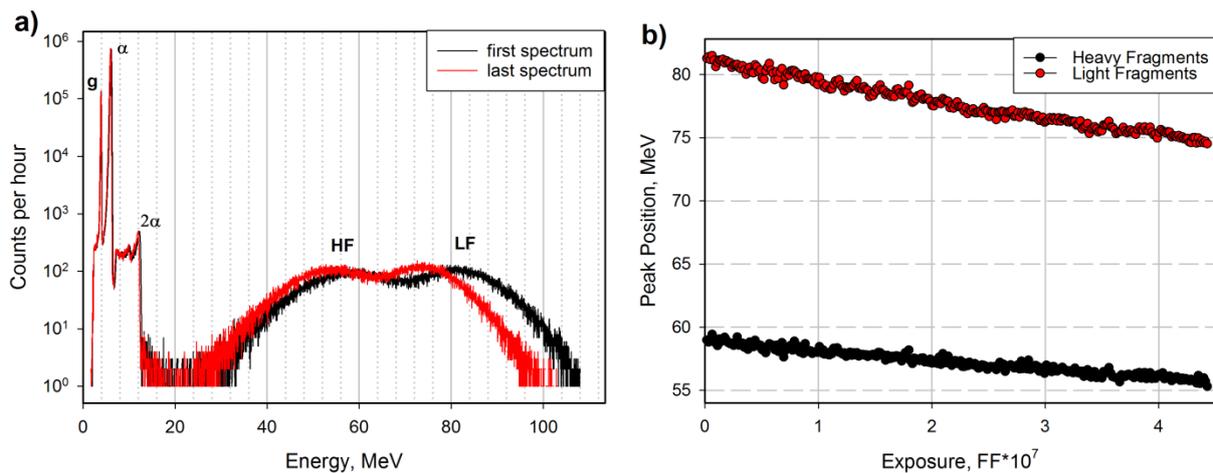

**Figure 1.** (a) The first and the last spectra measured by SB2 detector in the beginning and at the end of the prolonged irradiation period. The following peaks are marked: constant amplitude generator peak (g), peak of α-particles at 6.118 MeV (α), peak at doubled energy of α-particles (2α) and the peaks due to FFs of light (LF) and heavy (HF) groups. (b) Dependence of the light and heavy FF peaks visible energies on exposure by FFs.

## 3. Experimental results

In order to study the influence of temperature of irradiation on the degradation of the detector's parameters, the irradiation of identical SiLi detectors was performed at room (SiLi1 detector) and liquid nitrogen (SiLi2 detector) temperature, respectively. To study the influence of external electric field strength on the detector's parameters degradation, two identical SB detectors were subjected to the irradiation with different applied reverse biases, i.e. with different electric field strengths in their active regions. The operating biases applied to the respective detector during the irradiation period, the corresponding surface electric field strengths and the total exposures are collected in Table 1.

For all investigated detectors the similar signs of operational parameters degradation as a result of the prolonged irradiation by $^{252}$Cf fission products were revealed. As an example, Figure 1a represents the spectra recorded by SB2 detector at the beginning and at the end of the prolonged irradiation period. The peak at 6.1 MeV corresponding to α-particles and another peak at doubled energy of the α-particles caused by their accidental coincidences were used as reference points for the calibration of the energy scale. Two broad unresolved peaks appearing at higher energies correspond to fission fragments of light and heavy groups, respectively.

The main effect of the detector degradation is a gradual shift of fission fragments visible energy towards the lower values, see Figure 1a. The positions of the peaks corresponding to heavy (HF) and light (LF) fission fragments were approximated using the Gaussian function for each 1-hour series. The dependences of the peaks positions with exposure by fission fragments can be well described by linear functions (Figure 1b) for any masses of fission fragments and for all investigated detectors. The obtained slope coefficients are summarized in Table 1. It is interesting to note, that the obtained coefficients for the peaks of light and heavy fission fragments groups differ approximately by the factor of 2 – this holds for all types of investigated detectors and for all irradiation conditions. In more details this fact will be discussed separately in the next paper. A similar approximation of the positions of α-peaks didn't reveal any measurable shift with the irradiation dose for all studied detectors [4-5].

Another sign of the detector's operational parameters degradation under irradiation is the rapid increase of the leakage current which proceeds linear with the number of absorbed fission products [3]. The obtained slope coefficients of the leakage current growth are also collected in Table 1.

## 4. Discussion

It could be noted in Figure 1 that the peak energies of light and heavy groups of fission fragments are below the predicted values of 104 MeV and 79 MeV [10], respectively, even on the spectrum measured by non-irradiated detectors. The same is true for all other investigated detectors. This effect is known as pulse-height defect (PHD) in heavy charged particles spectroscopy by semiconductor detectors implying that the measured pulse height amplitude for heavy charged particles is somewhat lower than that for α-particles of the same energy [1]. It is generally considered that PHD is caused by a combination of energy losses (i) in the detector dead layer/entrance window, (ii) due to the atomic collisions and (iii) due to recombination of the electron-hole pairs created by the incident heavy particle. Whereas energy losses by (i) and (ii) mechanisms are well understood, the full understanding of the charge losses due to recombination is still missing. Two models were suggested supposing that enhanced carrier recombination proceeds either in the bulk region on the radiation-induced defects created by incident FFs [11], or at the surface states of the semiconductor [12]. The later model is consistent with the TRIM [13] simulation results (Figure 2) showing that the density of electron-hole pairs generated by fission fragments reaches the maximum in the near-surface region of the detector and then gradually drops down towards the bulk, suggesting therefore that decisive influence on PHD would have the carrier recombination at the surface states.

Previously, the PHD of about 7-10 MeV was reported for $^{252}$Cf fission fragments detection by semiconductor detectors not subjected to the prolonged irradiation [10]. These PHD values are close to those ones obtained for the investigated planar and SB1 detectors operated at high reverse bias – see Table 1. We believe, that higher PHD values in non-irradiated SiLi are related with rather thick entrance window in these detectors. Whereas the increase of PHD for SB2 detector operated at lower electric field (Table 1) reflects the influence of the electric field strength on the charge carrier collection efficiency, i.e. on the recombination of the generated electron-hole pairs (note that the active layer thickness in SB2 detector exceeds the projection range of incident FFs even at 30V).

**Table 1.** Irradiation conditions and the degradation of the operational parameters of the investigated detectors: $U_b$ – applied bias during irradiation; $F_s$ – surface electric field strength; $PHD_{LF}$/ $PHD_{HF}$ – pulse-height defects for light and heavy fragments peaks registered by non-irradiated detectors; $N_{FF}$ and $N_\alpha$ – exposure by fission fragments and α-particles, respectively; $\Delta E_{HF}/\Delta N_{FF}$ – slope coefficient describing the linear shift of heavy fission fragment maximum; $\Delta E_{LF}/\Delta N_{FF}$ – slope coefficient describing the linear shift of light fission fragment maximum; $\Delta I/\Delta N$ – rate of the reverse current increase relative to the total number of the registered fission products (wasn't measured for SiLi2 detector); $N_{FFmax}$ – maximal permissible exposure by fission fragments; $t$ – expected active operation period of the detector in a neutron source.

|  | p⁺n planar | SB1 | SB2 | SiLi1 | SiLi2 |
|---|---|---|---|---|---|
| $U_b$, V | 150 | 200 | 30 | 400 | 400 |
| $F_s$, kV/cm | 8.5 | 40 | 17 | 1.5 | 1.5 |
| $PHD_{LF}/PHD_{HF}$, MeV | 8/10 | 9/11 | 18/19 | 28/29 | 35/37 |
| $N_{FF}$ *$10^8$ | 1.1 | 0.45 | 0.43 | 3.4 | 1 |
| $N_\alpha$ *$10^{10}$ | 0.5 | 0.20 | 0.19 | 1.5 | 0.44 |
| $\Delta E_{HF}/\Delta N_{FF}$*$10^{-5}$, keV/FF | -0.9 | -1.8 | -8.9 | -3.6 | -5.7 |
| $\Delta E_{LF}/\Delta N_{FF}$*$10^{-5}$, keV/FF | -1.9 | -3.9 | -20 | -6.2 | -12 |
| $\Delta I/\Delta N$*$10^{-16}$, A/ion | 8.9 | 14 | 8.0 | 4.4 | - |
| $N_{FFmax}$ *$10^8$ | 22 | 12 | 2.2 | 6.9 | 4.7 |
| t, years | 11.6 | 6.3 | 1.2 | 3.6 | 2.5 |

As a result of the prolonged irradiation by $^{252}$Cf fission products, the linear shift of FF peaks positions, i.e. the linear increase of PHD for fission fragments peaks, was revealed. Since the task of semiconductor detector operating as a part of neutron calibration source is the reliable detection of fission fragments signal, the irradiated detector could be considered to be "degraded" when the spectrum of the heavy fission fragment overlaps with much more intense signal at double energy of α-peak, what prevents us from discrimination between them [3]. The values of maximal "permissible" exposure by fission fragments $N_{FFmax}$ corresponding to the beginning of the peaks overlap at three standard deviations from their maxima were estimated for each detector using the corresponding slope coefficients derived for HF peak and the results are presented in Table 1.

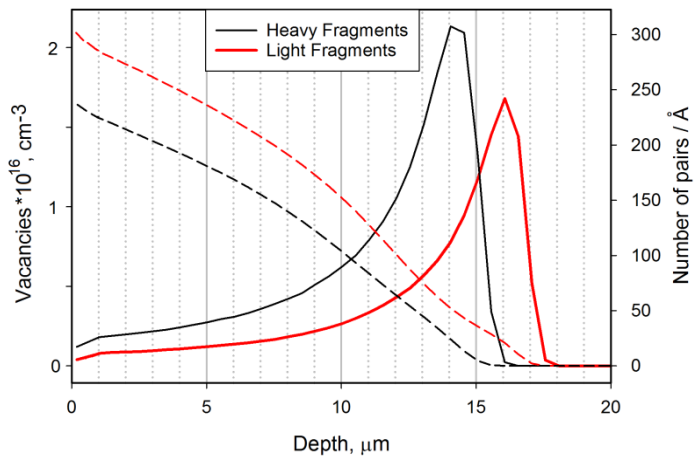

**Figure 2.** TRIM simulated vacancies distribution profiles (solid lines) and linear densities of electron-hole pairs (dashed lines) generated by light and heavy FFs with mean energies and masses of 104 MeV and 79 MeV, 106 amu and 142 amu, respectively.

Permissible exposure values $N_{FFmax}$ for the investigated detectors appeared to vary approximately by one order of magnitude. The highest $N_{FFmax}$ values were found for planar and SB1 detector operated at 200V. Reduction of the operating bias and thus the electric field strength in case of SB2 detector has led to considerable decrease of the expected permissible exposure value. Therefore, the electric field strength affects not only the PHD on non-irradiated detector, but also the value of the expected maximal exposure. However the $N_{FFmax}$ exposure values for SiLi detectors – which operated with lowest electric field as compared with other detectors – are significantly higher than that for SB2 detector. Thus the expected maximal exposure appeared to be more sensitive to the electric field strength in the surface barrier detectors and less sensitive in SiLi and planar detectors. It follows then that not only the electric field strength, but also a detector's internal structure defines the PHD growth under irradiation and the maximal permissible exposure.

According to TRIM simulations, irradiation of Si detectors by fission fragments will lead to the creation of vacancy-interstitial pairs and therefore to the formation of high density of radiation-induced defects in the region from detector surface till the depth of 17 μm with the maxima at 14-16 μm (Figure 2). Additionally TRIM indicates, that the energy of FFs is high enough to damage the detector surface by sputtering. Therefore, prolonged irradiation with fission fragments will lead to an increase of the carrier recombination rate both in Si bulk and on the surface of the semiconductor, thus contributing to the PHD growth.

The transition region in the detectors produced by planar and by SiLi technology ($p^+n$ and $p$-$i$ transition regions, respectively) is located inside the crystalline matrix at the typical depths of 50-500 nm from the surface. Apparently, the contribution of the surface recombination to the charge carrier losses will be more significant for surface-barrier detectors than for SiLi and planar ones, whereas the contribution of bulk defects – approximately similar in all detectors, what may be the reason for different sensitivity of $N_{FFmax}$ exposure to the electric field strength in these detectors. Additional investigations are needed to determine the dominant charge loss channel.

Suggested neutron calibration source should operate also at cryogenic temperatures (liquid nitrogen or slightly above). Performed irradiation of SiLi2 detector at liquid nitrogen temperature has shown, that in contrast to the electric field, temperature of irradiation seems to have no or only minor influence on the expected value of maximal exposure as it could be concluded from the comparable $N_{FFmax}$ values obtained for SiLi detectors irradiated at different temperatures. Somewhat smaller $N_{FFmax}$ exposure obtained for SiLi2 detector is probably related with thicker entrance window in this detector.

Knowing the maximal expected exposure values $N_{FFmax}$, it is possible to estimate the duration of active "lifetime" of neutron calibration source. For the operation of neutron calibration source the reasonable neutron activity would be the around 20 neutrons/s and taking into account that each spontaneous fission releases in average 3.7 fast neutrons, the activity of 20 neutrons/s would correspond to ~6 spontaneous fissions per second. Therefore, considering the maximal exposure value from Table 1, the duration of active "lifetime" of such neutron calibration source will be 1.2-11.6 years (without taking into account the decay of the radiation source).

During this operation period, a significant increase of leakage current up to ~100 μA can be expected at room temperature, as can be calculated from the obtained coefficients of leakage current growth (Table 1). Such high reverse current is unacceptable and therefore the detector cooling in order to reduce the reverse current during the neutron source operation will be required. The coefficients of current growth upon irradiation by fission products of $^{252}$Cf appeared to be an order of magnitude higher than the corresponding coefficients of 7-17×10$^{-17}$ A/α determined by us earlier for the identical detectors subjected to long-term irradiation by α-particles [4]. This fact confirms that prolonged irradiation by FFs leads to the creation of the effective recombination-generation defect centers participating in charge carrier recombination and the reverse current growth.

## 5. Conclusions
Prolonged irradiation of three different types of Si semiconductor detectors by fission products of $^{252}$Cf nuclide has led to a gradual increase of pulse-height defect for the fission fragments peaks in all investigated detectors. This will eventually lead to the overlap with more intense α-peak and therefore to the impossibility of further reliable detection of fission fragments by the semiconductor detector and thus to the limitation of the operation period of neutron calibration source. Obtained experimental results suggest, that in order to assure the longest operation period of the neutron calibration source it is worth to use the semiconductor detectors with lowest surface recombination rate and with highest possible electric field strength in their active region. Among the investigated detectors, the planar one most fully meets these requirements, whereas in relatively thick SiLi detectors it is difficult to achieve the high electric field strength and surface-barrier detectors may suffer from high surface recombination. With properly chosen semiconductor detector the expected active operation period of $^{252}$Cf-based neutron calibration source may reach up to 12 years.

**Acknowledgements**
The reported study was funded by RFBR, project number 20-02-00571